# Disclosing Generative AI Use in Digital Humanities Research

Rongqian Ma[1], Xuhan Zhang[2], Adrian Wisnicki[3]

[1]Indiana University Bloomington
[2]University of Missouri
[3]University of Nebraska-Lincoln

**ABSTRACT**

This survey study investigates how digital humanists perceive and approach generative AI (GenAI) disclosure in research. The results indicate that while digital humanities scholars acknowledge the importance of disclosing GenAI use, the actual rate of disclosure in research practice remains low. Respondents differ in their views on which activities most require disclosure and on the most appropriate methods for doing so. Most also believe that safeguards for AI disclosure should be established through institutional policies rather than left to individual decisions. The study's findings will offer empirical guidance to scholars, institutional leaders, funders, and other stakeholders responsible for shaping effective disclosure policies.

**INTRODUCTION**

Generative AI (GenAI) now permeates almost every aspect of daily life and has become integral to academic work. Although its use offers clear benefits to researchers, it also raises challenging questions about academic integrity and research ethics (Van Noorden et al., 2023; Hosseini et al., 2023; Abdelhafiz et al., 2024; Dedema & Ma, 2024; Ng et al., 2025). As a result, scholars continue to debate how best to acknowledge and attribute AI involvement in both professional and broader contexts.

Disclosure has therefore emerged as a potential policy solution. For example, Resnik et al. (2025) proposed a three-tier framework distinguishing mandatory, optional, and unnecessary disclosure, mirroring the increasingly nuanced guidance now issued by publishers. Publishers increasingly agree that GenAI tools cannot be listed as authors, yet their disclosure policies vary markedly. Most now distinguish between routine language polishing—grammar, spelling, formatting—and substantive content generation: the former typically escapes disclosure, whereas the latter demands it. For example, the STM (2023) association advised that authors need not declare AI-assisted copy-editing but must disclose any use that goes beyond those tasks. Implementation of AI disclosure principles at the journal level also varies. At the strict end, journals such as Wiley (2023) titles and *Science* (2023) require detailed statements describing the software, its provider, and even the prompts used, with authors affirming full responsibility for AI-generated text. *Springer Nature* (2023) adopts a moderate stance, insisting that generative use be "properly documented" while exempting minor editing. IOP Publishing (n.d.), acknowledging the ubiquity of these tools, merely encourages transparency and reminds authors they remain accountable for all content.

However, policies that mandate AI disclosure can produce unintended consequences. Empirical research has identified an "AI disclosure penalty" that surfaces in both professional spheres, such as news, business, healthcare, and personal or creative domains, including art, poetry, or mental health. The magnitude of the penalty varies by context: In professional settings, studies report moderate penalties (e.g., Reis et al., 2024; Proksch et al., 2024); while in creative contexts, penalties are generally smaller but still statistically significant (Horton Jr. et al., 2023). Studies have also consistently suggested the AI disclosure reduces perceived credibility across news, health, and business contexts, to name a few (Longoni et al., 2021; Henestrosa and Kimmerle, 2024; Reis et al., 2024).

Given these context-dependent penalties, understanding how different audiences interpret AI disclosure approach is important. Against this backdrop, some researchers adopt a community-centered perspective, examining stakeholders' needs and expectations around AI disclosure. A recent *Nature* survey, for example, reveals deep divisions among scholars over whether GenAI use should be declared at all and, if so, how much detail such declarations should include—from brief acknowledgements to full prompt disclosure (Kwon, 2025). We content that a community-centered lens is essential for navigating the complex terrain of AI disclosure, where formal policies intersect with the professional cultures that ultimately determine their uptake. Therefore, we adopt a community-based approach in this study, exploring how researchers view GenAI disclosure and how their views might shape future disclosure policies. We focus on the digital humanities community—an inherently interdisciplinary field that brings together scholars with diverse expertise (Luhmann & Burghardt, 2022)—to gauge attitudes toward declaring GenAI use in research. Specifically, we examine:

1) To what extent do digital humanists regard disclosure of GenAI use as necessary? And in what forms?
2) In which research contexts and for which research activities should GenAI involvement be disclosed?
3) Who should be responsible for developing and enforcing GenAI disclosure policies and practices?

By addressing these questions, we aim to capture the digital humanities community's collective views on GenAI disclosure and to advance discussions about field-specific effective strategies for incorporating AI responsively into research.

**METHODS**

We developed a Qualtrics survey to investigate these questions.[1] The survey contains 30 questions organized into 3 sections: (1) demographic background, (2) attitudes and preferred practices for GenAI disclosure, and (3) self-reported AI literacy. Due to the lack of a clear definition for "digital humanist" (Ma, 2022; Ramsay, 2011), we employed a self-identification approach: everyone who consider themselves as a digital humanities scholar was eligible to participate in the survey. The survey was distributed from 20 February to 6 May 2025 via social-media platforms (X, Bluesky, Slack, LinkedIn) and leading digital humanities mailing lists. We received 152 responses; after data cleaning, 99 fully completed surveys remained for analysis. Respondents span a wide range of DH subfields—literary and cultural studies, history, information science—and represent a geographically diverse community across the United States, United Kingdom, European Union, China, and beyond. An exploratory data analysis revealed the following initial insights.

**PRELIMINARY FINDINGS**

The findings show that 63% of the digital humanists (n=63) think it is necessary to disclose GenAI use. Among these respondents, the majority (54%, n=53) vote for "brief disclosure (model/version, parts generated by AI, etc.)" followed by "detailed disclosure (full description of process, example prompts, etc.)" (36%, n=36), "minimal disclosure (Yes/No)" (7%, n=7), and "prefer not to answer" (3%, n=3). Despite broad agreement on the value of transparency, practice of GenAI disclosure remains limited. Only 28% of the participants (n=28) who acknowledge the necessity of GenAI disclosure have actually disclosed in their research before. Among those who have disclosed, 20 report they have done so only 1-2 times, and they disclosed voluntarily (n=20). This gap shows that many researchers recognize the importance of disclosure, but they may lack the motivations or feel no external obligations to disclose in practice. The potential AI disclosure penalty might also account for this discrepancy between perception and practice.

---

[1] The survey is available at: https://docs.google.com/document/d/1OtcJzx5K-eSUG0yes-eYFU1wn2hSXh4x/edit?usp=sharing&ouid=112918067730057086842&rtpof=true&sd=true

The survey also indicates that digital humanists' disclosure priorities differ by activity and research context. Across the sample, the three activities most frequently prioritized for disclosure are (1) writing and rewriting, (2) data collection, transcription, translation, and analysis, and (3) peer review. Cross-tabulating this question (Q15) with respondents' primary research fields (Q5) reveals a largely consistent pattern: every discipline except computer science ranks "writing and rewriting" as the foremost activity that requires disclosure. Computer scientists instead place "brainstorming" at the top. Specifically, among respondents who rank "writing and rewriting" as the top activity requiring disclosure, 31 identify themselves as "Literary and Cultural Studies" scholars, highlighting the essential role of writing in their scholarship and the perceived obligation to disclose any GenAI contribution to that process.

Finally, we also find that most digital humanists think AI disclosure is an institutional, rather than individual, responsibility. Among the 99 surveyed participants, most distribute responsibility across the scholarly ecosystem, but with a clear hierarchy. Journal publishers and editorial boards are viewed as the primary gatekeepers (87%, n=86). Academic institutions follow closely (86 %, n=85), reflecting expectations that universities set and police their own standards. A second tier—professional societies (n=69), funding bodies (n=68), and ethical-review boards (n=61)—also commands majority support, signaling a preference for multilayered oversight within the research community itself. Government agencies receive moderate support (n=52), but confidence declines for actors more distant from day-to-day scholarship: industry organizations and independent oversight bodies (each n=37) and international entities such as UNESCO (n=36).

**CONCLUSION**

The survey study reveals a shared recognition of the importance of disclosing GenAI use, but divergent views on how such disclosure should be integrated into research practices. Next, we will examine whether factors such as scholars' AI literacy, personal AI experience, disciplinary affiliation, career stage, and seniority shape these views. The goal is to produce a comprehensive account of digital humanists' disclosure practices and attitudes. Over the longer term, we will administer a follow-up survey to track how digital humanists' collective sentiment evolves as AI technologies continue to advance.